\documentclass[aps,epsf,twocolumn,showpacs]{revtex4}
\usepackage{amsmath}
\usepackage{epsfig}

\begin{document}

\title{Critical behavior of the pure and random-bond two-dimensional triangular Ising ferromagnet}

\author{Nikolaos G. Fytas}\affiliation{Department of Physics, Section of Solid State
Physics, University of Athens, Panepistimiopolis, GR 15784
Zografos, Athens, Greece}

\author{Anastasios Malakis}
\affiliation{Department of Physics, Section of Solid State
Physics, University of Athens, Panepistimiopolis, GR 15784
Zografos, Athens, Greece}

\date{\today}

\begin{abstract}
We investigate the effects of quenched bond randomness on the
critical properties of the two-dimensional ferromagnetic Ising
model embedded in a triangular lattice. The system is studied in
both the pure and disordered versions by the same efficient
two-stage Wang-Landau method. In the first part of our study we
present the finite-size scaling behavior of the pure model, for
which we calculate the critical amplitude of the specific heat's
logarithmic expansion. For the disordered system, the numerical
data and the relevant detailed finite-size scaling analysis along
the lines of the two well-known scenarios - logarithmic
corrections versus weak universality - strongly support the
field-theoretically predicted scenario of logarithmic corrections.
A particular interest is paid to the sample-to-sample fluctuations
of the random model and their scaling behavior that are used as a
successful alternative approach to criticality.
\end{abstract}

\pacs{75.10.Nr, 05.50.+q, 64.60.Cn, 75.10.Hk} \maketitle

\section{Introduction}
\label{sec:1}

Understanding the role played by impurities on the nature of phase
transitions is of great importance, both from experimental and
theoretical perspectives. First-order phase transitions are known
to be dramatically softened under the presence of quenched
randomness~\cite{aizenman-89,hui-89,berker-93,chen-92,cardy-96,cardy-97,chatelain-98,paredes-99,chatelain-01,fernandez-08},
while continuous transitions may have their exponents altered
under random fields or random bonds~\cite{harris-74,chayes-86}.
There are some very useful phenomenological arguments and some,
perturbative in nature, theoretical results, pertaining to the
occurrence and nature of phase transitions under the presence of
quenched
randomness~\cite{hui-89,cardy-96,dotsenko-95,jacobsen-98}.
Historically, the most celebrated criterion is that suggested by
Harris~\cite{harris-74}. This criterion relates directly the
persistence, under random bonds, of the non random behavior to the
specific heat exponent $\alpha_{p}$ of the pure system. According
to this criterion, if $\alpha_{p}>0$, then disorder will be
relevant, i.e., under the effect of the disorder, the system will
reach a new critical behavior. Otherwise, if $\alpha_{p}<0$,
disorder is irrelevant and the critical behavior will not change.

Pure systems with a zero specific heat exponent ($\alpha_{p}=0$)
are marginal cases of the Harris criterion and their study, upon
the introduction of disorder, has been of particular
interest~\cite{MK-99}. The paradigmatic model of the marginal case
is, of course, the general random 2d Ising model (random-site,
random-bond, and bond-diluted) and this model has been extensively
investigated and debated [see Ref.~\cite{gordillo-09} and
references therein]. Several recent studies, both analytical
(renormalization group and conformal field theories) and numerical
(mainly Monte Carlo (MC) simulations) devoted to this model, have
provided very strong evidence in favor of the so-called
logarithmic corrections's
scenario~\cite{DD-81,jug-83,shalaev-84,shankar-87,ludwig-87}.
According to this, the effect of infinitesimal disorder gives rise
to a marginal irrelevance of randomness and besides logarithmic
corrections, the critical exponents maintain their 2d Ising
values. In particular, the specific heat is expected to slowly
diverge with a double-logarithmic dependence of the form $C
\propto \ln (\ln t)$, where $t=|T-T_{c}|/T_{c}$ is the reduced
critical
temperature~\cite{DD-81,jug-83,shalaev-84,shankar-87,ludwig-87}.
Here, we should mention that there is not full agreement in the
literature and a different scenario, the so-called weak
universality scenario~\cite{KP-94,kuhn-94,suzuki-74,gunton-75},
predicts that critical quantities, such as the magnetization,
zero-field susceptibility, and correlation length display
power-law singularities, with the corresponding exponents $\beta$,
$\gamma$, and $\nu$ changing continuously with the disorder
strength; however this variation is such that the ratios
$\beta/\nu$ and $\gamma/\nu$ remain constant at the pure system's
value. The specific heat of the disordered system is, in this
case, expected to saturate, with a corresponding correlation
length's exponent $\nu\geq 2/d$~\cite{chayes-86}.

In general, a unitary and rigorous physical description of
critical phenomena in disordered systems still lacks and
certainly, lacking such a description, the study of further models
for which there is a general agreement in the behavior of the
corresponding pure cases is very important. In this sense, the
triangular Ising ferromagnet is a further suitable candidate for
testing the above predictions that has not been previously
investigated in the literature. Thus, our investigation will be
related to the extensive relevant literature concerning the
critical properties of the disordered 2d square Ising
model~\cite{MK-99,gordillo-09,DD-81,jug-83,shalaev-84,shankar-87,ludwig-87,KP-94,kuhn-94,LC-87,mayer-89,wang-90,ludwig-90,
ziegler-90,heuer-92,shalaev-94,QS-94,TS-94,
MS-95,AQS-96,JS-96,CHMP-97,BFMMPR-97,AQS-97,SSLI-97,
RAJ-98,SSV-98,AQS-99,LSZ-01,nobre-01,SV-01,TO-01,MC-02,
COPS-04,queiroz-06,LQ-06,PHP-06,kenna-06,MP-07,hasenbusch-08,hadjiagapiou-08,kenna-08,fytas-08a}.
In particular, our discussion will focus on the main point of
dispute over the last two decades, concerning the two well-known
conflicting scenarios mentioned above and we will provide
additional new evidence in favor of the well-established scenario
of strong universality. We should note here that, the
theoretically predicted scenario of strong universality has been
confirmed by several MC studies on the square lattice starting
from the early 90's to nowadays [see Ref.~\cite{gordillo-09} for a
detailed historical review].

As mentioned above, it is always important to consider further
models for which the critical properties of the corresponding pure
versions are exactly known. This is also the case for the present
model under consideration, namely the triangular Ising model,
called hereafter as the TrIM, defined as usual by the Hamiltonian
\begin{equation}
\label{eq:1} H=-J\sum_{<ij>}s_{i}s_{j},
\end{equation}
where the spin variables $s_{i}$ take on the values $-1,+1$,
$<ij>$ indicates summation over all nearest-neighbor pairs of
sites, and $J>0$ is the ferromagnetic exchange interaction. The
TrIM belongs to the same universality class with the corresponding
square Ising model (SqIM), sharing the same values of critical
exponents and a logarithmic behavior of the specific
heat~\cite{fisher-67,mon-93}. Additionally, the critical
temperature of the model and also the critical amplitude $A_{0}$
of Ferdinand and Fisher's~\cite{ferdinand-69} specific heat's
logarithmic expansion [see also the discussion in Sec.~\ref{sec:3}
and Eq.~(\ref{eq:7})] are exactly known from the early work of
Houtappel~\cite{houtappel-50} to be $T_{c}=4/\ln {3}=3.6409\cdots$
and $A_{0}=0.499069\cdots$, respectively. Nevertheless, it appears
that for the TrIM a verification of the finite-size scaling (FSS)
properties of the model using high quality data from MC simulation
is still lacking. Thus, the first part of this work is devoted to
the investigation of the FSS behavior of the model, especially the
estimation of the amplitudes and other relevant coefficients in the specific heat's
logarithmic expansion and also to the estimation of the critical
exponents. In this sense, the aim of this first part is twofold:
First, to provide the first detailed FSS analysis of the pure
model and, second, to present a concrete reliability test of the
proposed numerical scheme.

Our main focus, on the other hand, is the case with bond disorder
given by the bimodal distribution
\begin{align}
\label{eq:2}
P(J_{ij})~=~&\frac{1}{2}~[\delta(J_{ij}-J_{1})+\delta(J_{ij}-J_{2})]\;;\\
\nonumber
&\frac{J_{1}+J_{2}}{2}=1\;;\;\;J_{1}>J_{2}>0\;;\;\;r=\frac{J_{2}}{J_{1}}\;,
\end{align}
so that $r$ reflects the strength of the bond randomness and we
fix $2k_{B}/(J_{1}+J_{2})=1$ to set the temperature scale. The
value of the disorder strength considered throughout this paper is
$r=1/3$. The resulting quenched disordered
(random-bond) version of the Hamiltonian defined in
Eq.~(\ref{eq:1}) reads now as
\begin{equation}
\label{eq:3} H=-\sum_{<ij>}J_{ij}s_{i}s_{j}
\end{equation}
and will be referred in the sequel as the random-bond triangular
Ising model (RBTrIM). The corresponding random-bond SqIM will be
denoted hereafter respectively as RBSqIM. The model on the square
lattice has the advantage that the critical temperature is exactly
known as a function of the disorder strength $r$ by duality
relations~\cite{fisch-78}. For the RBTrIM there exist only several
approximations for the critical frontier of the site- and
bond-diluted cases, obtained via renormalization-group
techniques~\cite{yeomans-79} and, to our knowledge, a study of the
critical behavior of the model is lacking.

The rest of the paper is laid out as follows: In Sec.~\ref{sec:2}
we present the necessary simulation details of our numerical
scheme. In Sec.~\ref{sec:3} we discuss the FSS behavior of the
pure model, testing with our high accuracy numerical data the
exact expansion of the critical specific heat. Then, in
Sec.~\ref{sec:4} we present a detailed FSS analysis for the random
version of the model, including - apart from the classical FSS
techniques - concepts from the scaling theory of disordered
systems. Our results and the relevant discussion clearly favors
the scenario of strong universality in marginal disordered
systems. Finally, Sec.~\ref{sec:5} summarizes our conclusions.

\section{Simulation Details}
\label{sec:2}

Resorting to large scale MC simulations is often
necessary~\cite{selke-94}, especially for the study of the
critical behavior of disordered systems. It is also well
known~\cite{newman-99} that for such complex systems traditional
methods become very inefficient and that in the last few years
several sophisticated algorithms, some of them are based on
entropic iterative schemes, have been proven to be very effective.
The present numerical study of the RBTrIM will be carried out by
applying our recent and efficient entropic
scheme~\cite{fytas-08a,malakis-04,fytas-08b}. In this approach we
follow a two-stage strategy of a restricted entropic sampling,
which is described in our study of random-bond Ising models (RBIM)
in 2d~\cite{fytas-08a} and is very similar to the one applied also
in our numerical approach to the 3d random-field Ising model
(RFIM)~\cite{fytas-08b}. In these papers, we have presented in
detail the various sophisticated routes used for the restriction
of the energy subspace and the implementation of the Wang-Landau
(WL) algorithm~\cite{wang-01}. Further details and an up to date
implementation of this approach, especially for the study of
disordered systems, is provided in our recent paper on the
universality aspects of the pure and random-bond 2d Blume-Capel
model~\cite{malakis-10}.

We do not wish to reproduce here the details of our two-stage
implementation and the practice followed in our scheme for
improving accuracy by repeating the simulations. However, we
should like to include a brief discussion on the approximate
nature of the WL method. The usual WL recursion proceeds by
modifying the density of states $G(E)$ according to the rule
$G(E)\rightarrow f G(E)$ and initially one chooses $G(E)=1$ and
$f=f_{0}=e$. Once the accumulative energy histogram is
sufficiently flat, the modification factor $f$ is redefined as:
$f_{j+1}=\sqrt{f_{j}}$, with $j=0,1,\ldots$ and the energy
histogram reset to zero until $f$ is very close to unity (i.e.
$f=e^{10^{-8}}\approx 1.000\; 000\; 01$). As has been reported by
many authors in the study of several models, once $f$ is close
enough to unity, systematic deviations become negligible.
\begin{figure}[htbp]
\includegraphics*[width=8 cm]{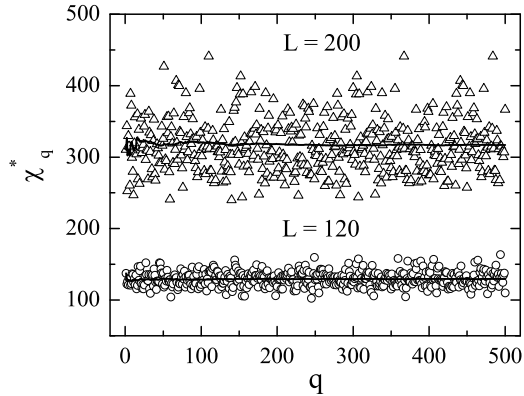}
\caption{\label{fig:1} Disorder distribution of the susceptibility
maxima of two lattices with linear sizes $L=120$ and $L=200$ of
the RBTrIM. The running averages over the samples is shown by the
thick solid lines.}
\end{figure}
However, the WL recursion violates the detailed balance from the
early stages of the process and care is necessary in setting up a
proper protocol of the recursion. In spite of the fact that the WL
method has produced very accurate results in several models, it is
fair to say that there is not a safe way to access possible
systematic deviations in the general case. This has been pointed
out and critiqued in a recent review by Janke~\cite{janke-08}. For
the 2d Ising model (where exact results are available for
judgement) the WL method has been shown to converge very rapidly.
Furthermore, from our experience and especially from our recent
study of the 2d RBSqIM~\cite{fytas-08a}, for which the exact phase
diagram is known by duality relations, our high-level WL
implementation has produced excellent results, enabling us to
discriminate between competing theoretical predictions on that
model. Since the RBTrIM is expected to have a similar ``entropy
structure'' to the corresponding square model, we anticipate and
it will be verified in the sequel, that, our WL scheme produces
sufficiently accurate estimates enabling us, also in this case, to
distinguish between competing theoretical predictions, as we have
already done in the corresponding model on the square lattice.

Using this scheme we performed extensive simulations for several
lattice sizes in the range $L=20-200$, over large ensembles
$\{1,\cdots,q,\cdots,Q\}$ of random realizations ($Q=500$). Let us
note here that for the pure model we simulated for each lattice
size, $200$ independent runs (WL random walks). It is well known
that, extensive disorder averaging is necessary when studying
random systems, where usually broad distributions are expected
leading to a strong violation of
self-averaging~\cite{aharony-96,wiseman-98}. A measure from the
scaling theory of disordered systems, whose limiting behavior is
directly related to the issue of
self-averaging~\cite{aharony-96,wiseman-98} may be defined with
the help of the relative variance of the sample-to-sample
fluctuations of any relevant singular extensive thermodynamic
property $Z$ as follows:
$R_{Z}=([Z^{2}]_{av}-[Z]^{2}_{av})/[Z]^{2}_{av}$.
Figure~\ref{fig:1} presents evidence that the above number of
random realizations is sufficient in order to obtain the true
average behavior and not a typical one. In particular, we plot in
this figure (for lattice sizes $L=120$ and $L=200$) the disorder
distribution of the susceptibility maxima $\chi_{q}^{\ast}$ and
the corresponding running average, i.e. a series of averages of
different subsets of the full data set - each of which is the
average of the corresponding subset of a larger set of data
points, over the samples for the simulated ensemble of $Q=500$
disorder realizations. A first striking observation from this
figure is the existence of very large variance of the values of
$\chi_{q}^{\ast}$, indicating the expected violation of
self-averaging for this quantity. This figure illustrates that the
simulated number of random realizations is sufficient in order to
probe correctly the average behavior of the system, since already
for $Q\approx 300$ the average value of $\chi^{\ast}_{q}$ appears
quite stable.

Closely related to the above issue of self-averaging in disordered
systems is the manner of averaging over the disorder. This
non-trivial process may be performed in two distinct ways when
identifying the finite-size anomalies, such as the peaks of the
magnetic susceptibility. The first way corresponds to the average
over disorder realizations ($[\ldots]_{av}$) and then taking the
maxima ($[\ldots]^{\ast}_{av}$), or taking the maxima in each
individual realization first, and then taking the average
($[\ldots^{\ast}]_{av}$). In the present paper we have undertaken
our FSS analysis using both ways of averaging and have found
comparable results for the values of the critical exponents, as
will be discussed in more detail below. Closing this brief
outline, let us comment on the statistical errors of our numerical
data. The statistical errors of our WL scheme on the observed
average behavior, were found to be of relatively small magnitude
(of the order of the symbol sizes) when compared to the relevant
disorder-sampling errors (due to the finite number of simulated
realizations). Thus, the error bars in most of our figures below
concerning the average $[\ldots]^{\ast}_{av}$ and used also in the
corresponding fitting attempts, reflect the disorder-sampling
errors and have been estimated using groups of $50$ realizations
via the jackknife method~\cite{newman-99}. On the other hand for
the case $[\ldots^{\ast}]_{av}$ the error bars shown reflect the
sample-to-sample fluctuations.

\section{Pure Model}
\label{sec:3}

In this Section we proceed to investigate the critical behavior of
the pure TrIM defined in Eq.~(\ref{eq:1}). Our aim is to observe
the exact critical behavior of the model and also to estimate the
whole set of critical exponents, paying particular attention to
the FSS behavior of the critical specific heat. As mentioned
above, the numerical data shown below in Figs.~\ref{fig:2} -
\ref{fig:4} have been estimated as averages over $200$ independent
runs together with the corresponding error bars.

\begin{figure}[htbp]
\includegraphics*[width=7 cm]{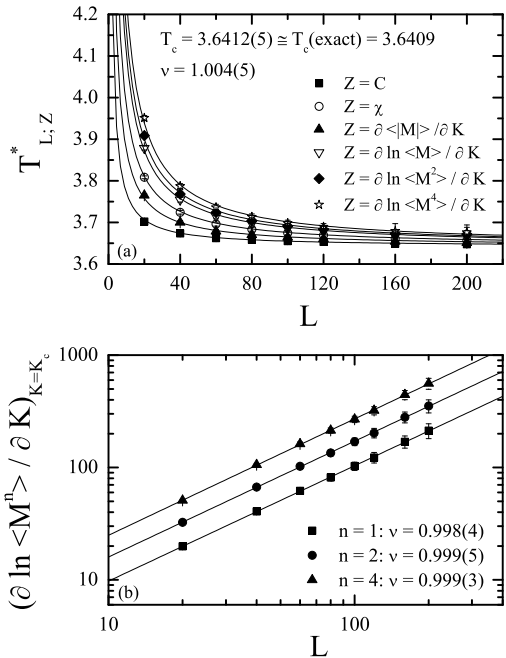}
\caption{\label{fig:2} (a) Simultaneous fitting of the
form~(\ref{eq:6}) of $6$ pseudocritical temperatures defined in
the text. (b) FSS of several powers of the logarithmic derivatives
[Eq.~(\ref{eq:5})] of the order parameter at the critical
temperature in a log-log scale. The solid lines are linear
fittings giving the value $\nu=1$ for the correlation length's
exponent.}
\end{figure}
Figure~\ref{fig:2}(a) gives the shift behavior of the
pseudocritical temperatures corresponding to the peaks of the
following six quantities: specific heat $C$, magnetic
susceptibility $\chi$, derivative of the absolute order parameter
with respect to inverse temperature $K=1/T$~\cite{ferrenberg-91}
\begin{equation}
\label{eq:4} \frac{\partial \langle |M|\rangle}{\partial
K}=\langle |M|H\rangle-\langle |M|\rangle\langle H\rangle,
\end{equation}
and logarithmic derivatives of the first ($n=1$), second ($n=2$),
and fourth ($n=4$) powers of the order parameter with respect to
inverse temperature~\cite{ferrenberg-91}
\begin{equation}
\label{eq:5} \frac{\partial \ln \langle M^{n}\rangle}{\partial
K}=\frac{\langle M^{n}H\rangle}{\langle M^{n}\rangle}-\langle
H\rangle.
\end{equation}
Fitting our data for the whole lattice range to the expected
power-law behavior
\begin{equation}
\label{eq:6} T^{\ast}_{L;Z}=T_{c}+bL^{-1/\nu},
\end{equation}
where $Z$ stands for the different thermodynamic quantities
mentioned above, we find the critical temperature to be
$T_{c}=3.6412(5)$ which is in excellent agreement with the exact
value $3.6409\cdots$. Additionally, our estimate of the critical
exponent $\nu$ of the correlation length is $\nu=1.004(5)$, also
in excellent agreement with the value $\nu=1$.
\begin{figure}[htbp]
\includegraphics*[width=7 cm]{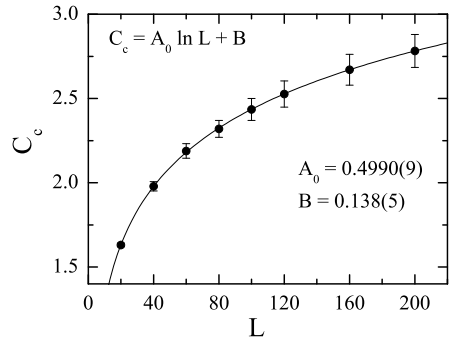}
\caption{\label{fig:3} FSS of the specific heat data at the exact
critical temperature.}
\end{figure}
Additional estimates for the critical exponent $\nu$ may be
obtained from the scaling behavior of the logarithmic
derivatives~(\ref{eq:5}), which scale as $\sim L^{1/\nu}$ with the
lattice size~\cite{ferrenberg-91}. The FSS of these logarithmic
derivatives at the critical temperature is shown in
Fig.~\ref{fig:2}(b) and in all three cases a value $\nu\approx 1$
is obtained consistent with the estimate from panel (a) of
Fig.~\ref{fig:2} and with the exact value $\nu=1$.

We know turn to the most interesting issue in the study of the pure model,
which is the specific heat's logarithmic expansion, as mentioned in the introduction.
For the square lattice, it was shown in 1969 in the pioneering work of Ferdinand and
Fisher~\cite{ferdinand-69} that close to the critical point the specific heat obeys the following
FSS expansion
\begin{equation}
\label{eq:7}
C_{L}(T)=A_{0}\ln{L}+B(T)+B_{1}(T)\frac{\ln{L}}{L}+B_{2}(T)\frac{1}{L}+\cdots,
\end{equation}
where the value of the critical amplitude $A_{0}$ is $0.494358\cdots$. As
pointed out in Ref.~\cite{ferdinand-69}, this is the same with the amplitude $A_{0}$
in the temperature expansion of the specific heat close to the critical point and this was already known
from the original paper of Onsager~\cite{onsager}. The first $B$ coefficients are given in
Ref.~\cite{ferdinand-69} and further details have been presented in
Refs.~\cite{malakis-04,janke-02,salas-01,wu-03}. In particular,
at the critical temperature the constant term $B$ is $0.138149\cdots$
and as it is also well-known~\cite{ferdinand-69,salas-01} the coefficient $B_{1}$ is zero.

The universality of the above expansion has been already pointed
out and discussed by Fisher~\cite{fisher-67}. For the three most
common 2d lattices, i.e. the square, plane triangular, and
honeycomb, a unified approach has been presented by Wu \emph{et
al.}~\cite{wu-03}, from which one can find also for the plane
triangular lattice the first $B$-coefficients and the amplitude
$A_{0}$ at the critical temperature.
\begin{figure}[htbp]
\includegraphics*[width=7 cm]{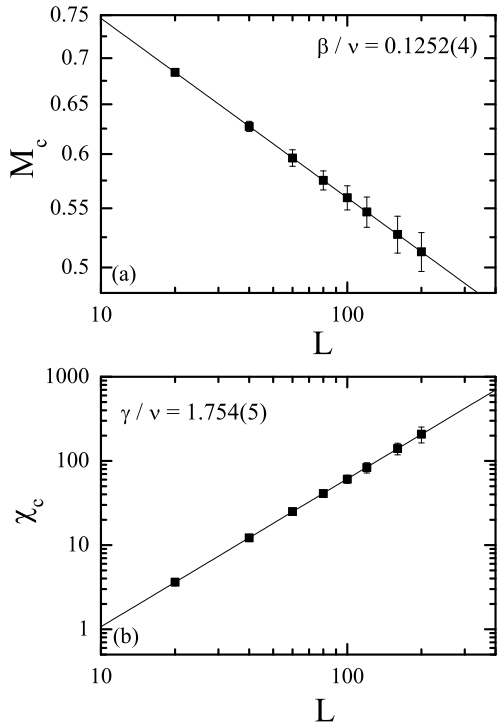}
\caption{\label{fig:4} Magnetic exponent ratios of the pure TrIM:
FSS in a log-log scale of (a) the order parameter and (b) the
magnetic susceptibility at the critical temperature. In both cases
linear fittings are applied.}
\end{figure}
As shown by Wu \emph{et al.}~\cite{wu-03} for all the three
lattices, the coefficient $B_{1}$ is zero at the critical
temperature and as in the square lattice, also for the triangular
lattice, the critical amplitude of the expansion is
$A_{0}=0.499069\cdots$, equal to the amplitude of the temperature
expansion close to the critical point obtained by
Houtappel~\cite{houtappel-50}. From the work of Wu \emph{et
al.}~\cite{wu-03} we find, using their closed form expressions,
for the plane triangular lattice (with periodic boundary
conditions and aspect ratio $R=1$) the following values
$B=B(T_{c})=0.14185\cdots$ and
$B_{2}=B_{2}(T_{c})=-0.15003\cdots$.

It is of interest at this point to examine the compatibility of
our numerical data with the above expansion for the case of the
critical specific heat data, i.e. the data of the specific heat at
the exact critical temperature of the TrIM. In Fig.~\ref{fig:3} we consider only
the first two terms in the above expansion (i.e. we set also $B_{2}=0$) and pay
attention in estimating the critical amplitude $A_{0}$ and the
constant $B$-contribution. From the results of the fitting in the
total lattice range $L=20-200$, as also shown in the figure, we
observe that the estimated value for the critical amplitude
$A_{0}=0.4990(9)$ is very close to the exact value
$A_{0}=0.499069\cdots$~\cite{houtappel-50}. For the first
$B$-coefficient we find the value $B=0.138(5)$ which is in good
agreement with the exact result $0.14185\cdots$. Let us note here that if we try to fit
the data of Fig.~\ref{fig:3} including also the coefficient
$B_{2}$ of the expansion~(\ref{eq:7}) we get the estimates
$0.4952(59)$, $0.153(16)$, and $-0.20(12)$ for the critical
amplitude $A_{0}$ and the coefficients $B$ and $B_{2}$,
respectively. However, if we fix the values of $A_{0}$ and $B$ to
their exact ones, the estimate for $B_{2}$ we get from the fitting
is $-0.158(11)$, in excellent agreement with the exact
value, whereas if we fix the value of $B$ and $B_{2}$ to their exact ones
we get the estimate $0.4990(3)$ for the critical amplitude $A_{0}$.

Finally, Figs.~\ref{fig:4}(a) and (b) present our estimations for
the magnetic exponent ratios $\beta/\nu$ and $\gamma/\nu$. For the
estimation of $\beta/\nu$ we use the values of the order parameter
at the exact critical temperature. As shown in panel (a), in a
log-log scale, the linear fitting provides the estimate
$\beta/\nu=0.1252(4)$. In panel (b) we show the FSS of the
critical susceptibility, also in a log-log scale. The straight
line is a linear fitting for $L\geq 20$ giving the estimate
$\gamma/\nu=1.754(5)$. Thus, our results presented in this Section
for the pure 2d TrIM model are in excellent agreement with the
exact results and also with the expected logarithmic expansion of
the specific heat~\cite{ferdinand-69,wu-03}. This consists a very
strong accuracy test of the proposed two-stage WL entropic
sampling in restricted energy spaces.

\section{Random Model}
\label{sec:4}

We now present our numerical results for the random-bond version
of the triangular Ising model for disorder strength
$r=1/3$. From simple universality-type theoretical
arguments, this system is also expected to undergo a second-order
transition between the ferromagnetic and paramagnetic phases and
in particular it should be also expected that this transition will be in the
same universality class as the RBSqIM.

\begin{figure}[htbp]
\includegraphics*[width=7 cm]{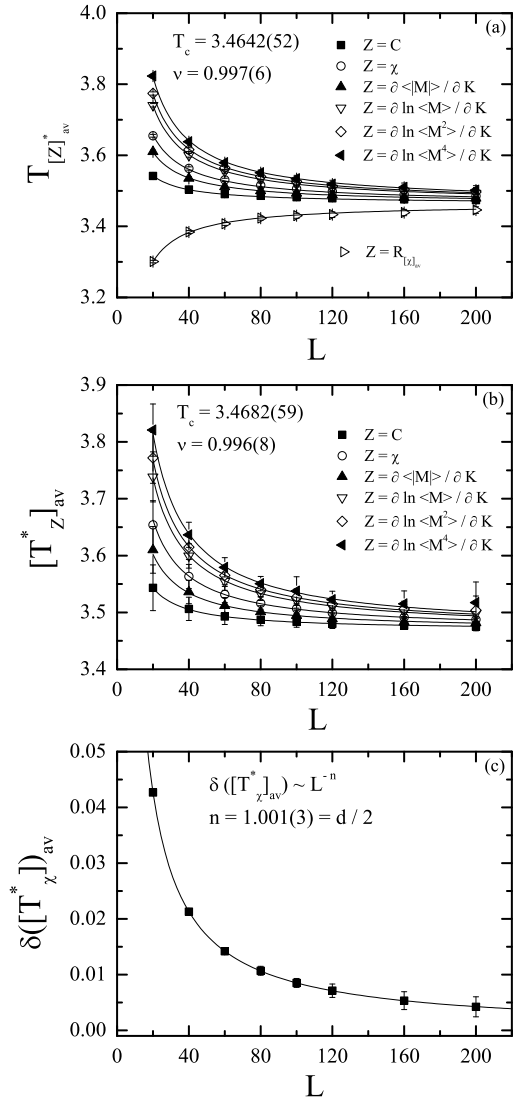}
\caption{\label{fig:5} (a) - (b) Shift behavior of several
pseudocritical temperatures defined in the text. The error bars in
panel (b) reflect the sample-to-sample fluctuations. (c) FSS of
the sample-to-sample fluctuations of the pseudocritical
temperature of the magnetic susceptibility shown in panel (b).}
\end{figure}

We start our analysis of the RBTrIM by presenting the general
shift behavior of various pseudocritical temperatures of the
model. Figure~\ref{fig:5}(a) illustrates the shift behavior of $7$
pseudocritical temperatures defined as the temperature where the
corresponding average thermodynamic property attains its maximum.
The first $6$ are as those defined in Sec.~\ref{sec:2} for the
corresponding pure model. The last pseudocritical temperature is a
newly introduced temperature, defined as the temperature where the
ratio $R_{[\chi^{\ast}]_{av}}$ defined in Sec.~\ref{sec:2},
becomes maximum. The solid lines show an excellent simultaneous
power-law fitting attempt of the form~(\ref{eq:6}) giving the
value $T_{c}=3.4642(52)$ for the critical temperature of the
random model and a value $\nu=0.997(6)$ for the critical exponent
$\nu$ of the correlation length. The fitting shown in
Fig.~\ref{fig:5}(a) has been performed for all lattice sizes and
it was very stable when shifting the $L$-range to larger values.
Note also that a simple fitting attempt using only the newly
defined pseudocritical temperature defined with the help of the
sample-to-sample fluctuations gives a value $T_{c}=3.4677(47)$ for
the critical temperature and a value $\nu=0.994(8)$ for the
correlation length's exponent. These overall estimates for the
exponent $\nu$ consist a strong indication that the RBTrIM shares
the same value of $\nu$ as the pure version, thus reinforcing the
scenario of logarithmic corrections (strong universality).

Figure~\ref{fig:5}(b) illustrates again the shift behavior of the
$6$ pseudocritical temperatures of panel (a), estimated now via
the second way of averaging discussed in Sec.~\ref{sec:1}, i.e. by
taking the average over the individual pseudocritical
temperatures. The error bars shown in this panel reflect the
sample-to-sample fluctuations of the pseudocritical temperatures.
Again, the results obtained from the simultaneous fitting attempt
$T_{c}=3.4682(59)$ and $\nu=0.996(8)$, as also shown in the
figure, agree excellently with the estimates of panel (a),
providing further evidence in favor of the accuracy of our
numerical scheme and the strong universality hypothesis.
\begin{figure}[htbp]
\includegraphics*[width=7 cm]{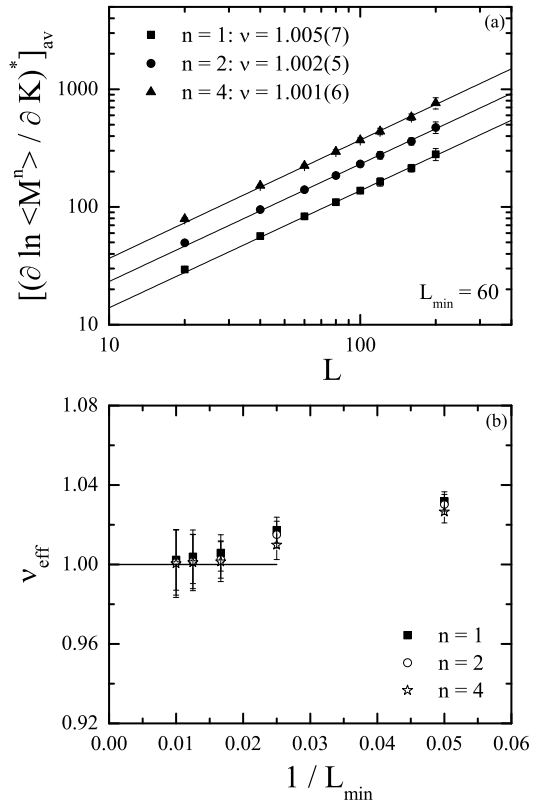}
\caption{\label{fig:6} (a) FSS of the logarithmic derivatives of
the order parameter in a log-log scale. The solid lines are simple
linear fittings for the larger lattice sizes ($L_{min}=60$). (b)
Values of effective exponents $\nu_{eff}$ obtained from the data
of panel (a) from several fitting attempts in the range
$(L_{min}-L_{max})$. The solid line marks the proposed estimate
$\nu=1$.}
\end{figure}
Noteworthy that, if we fix the exponent $\nu$ to the exact value
$\nu=1$ we get the most accurate estimates for the critical
temperature to be $3.4663(16)$ and $3.4669(19)$ from the
corresponding fittings of panels (a) and (b) of Fig.~\ref{fig:5}.

Using now the above sample-to-sample fluctuations of the
pseudocritical temperatures and the theory of FSS in disordered
systems as introduced by Aharony and Harris~\cite{aharony-96} and
Wiseman and Domany~\cite{wiseman-98}, one may further examine the
nature of the fixed point that controls the critical behavior of
the disordered system. According to the theoretical
predictions~\cite{aharony-96,wiseman-98}, the pseudocritical
temperatures $T_{Z}^{\ast}$ of the disordered system are
distributed with a width $\delta [T_{Z}^{\ast}]_{av}$, that scales
with the system size as
\begin{equation}
\label{eq:8} \delta([T_{Z}^{\ast}]_{av})\sim L^{-n},
\end{equation}
where $n=d/2$ or $n=1/\nu_{r}$, depending on whether the
disordered system is controlled by the pure or the random fixed
point, respectively. In panel (c) of Fig.~\ref{fig:5} we plot
these sample-to-sample fluctuations of the pseudocritical
temperature of the magnetic susceptibility.
\begin{figure}[htbp]
\includegraphics*[width=8 cm]{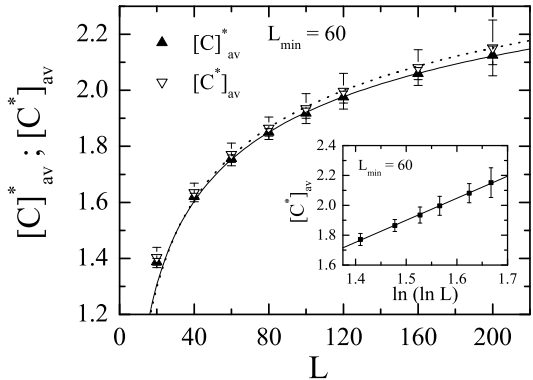}
\caption{\label{fig:7} FSS of the specific heat maxima obtained
via the two distinct ways of disorder averaging. The solid and
dotted lines are double-logarithmic fittings of the
form~(\ref{eq:9}) for lattice sizes in the range $L=60-200$. The
inset shows the data for the case $[C^{\ast}]_{av}$ as a function
of the double logarithm of $L$. The solid line is an excellent
linear fitting.}
\end{figure}
The solid line shows a very good power-law fitting giving the
value $1.001(3)$ for the exponent $n$ of the theory, which is in
agreement with the case $n=d/2$, indicating that the random model
is controlled by the pure fixed point. We should note here that
analogous results to those discussed here in panel (c) for the
case of the site-diluted Ising model on the square lattice have
been presented by Tomita and Okabe, using the probability-changing
cluster algorithm~\cite{TO-01}.

As in the pure case, the second alternative estimation of $\nu$ is
carried out by analyzing the divergency of the logarithmic
derivatives of the order parameter. In Fig.~\ref{fig:6}(a) we
illustrate in a double-logarithmic scale the size dependence of
the first- (filled squares), second- (filled circles), and
fourth-order (filled triangles) logarithmic derivatives (averaged
over the individual maxima). The solid lines show linear fittings
for the sizes $L\geq 60$. In all cases a value $\nu=1$ is obtained
for the critical exponent $\nu$, providing further evidence to the
strong universality scenario emerged from Fig.~\ref{fig:5}.
Figure~\ref{fig:6}(b) illustrates our method to evaluate and
discuss the stability of the estimation for the exponent $\nu$
from the scaling behavior of the logarithmic derivatives of panel
(a). It shows values of effective exponents ($\nu_{eff}$)
determined by imposing a lower cutoff $(L_{min})$ and applying
simultaneous fittings in windows $(L_{min}-L_{max})$, where as for
the pure case, $L_{max}=200$ and $L_{min}=20,40,60,80$, and $100$
as a function of $1/L_{min}$. The effective estimates show a
finite-size effect for small values of the lower cutoff, whereas
and for $L\geq 60$ a clear trend towards the value $\nu=1$ of the
Ising universality class is obtained. Let us note here that the
same picture emerged from the FSS of the disorder-averaged
logarithmic derivatives of the form $[\partial \ln {\langle
M^{n}}\rangle/\partial K]^{\ast}_{av}$ that corresponds to the
first way of averaging, but is omitted here for brevity. We should
note here that a similar cross-over behavior in the estimates of
the critical exponent $\nu$ has been observed in the case of the
2d site-diluted SqIM by Ballesteros \emph{et al.}~\cite{BFMMPR-97}
and has been explained as logarithmic corrections.
\begin{figure}[htbp]
\includegraphics*[width=7 cm]{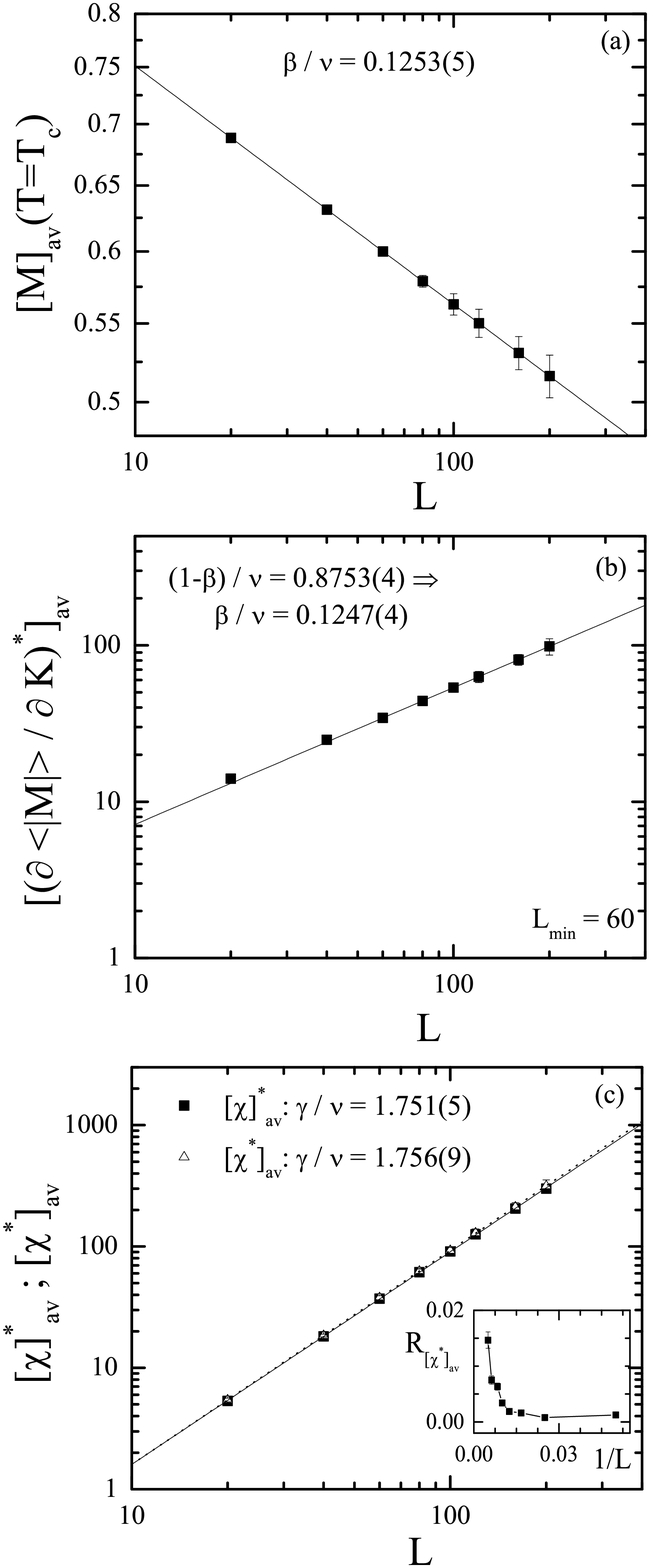}
\caption{\label{fig:8} (a) FSS of the average order parameter at
the critical temperature. (b) FSS of the disorder-averaged
inverse-temperature derivative of the absolute order parameter.
(c) FSS of the disorder-averaged magnetic susceptibility. The
inset shows the limiting behavior of the ratio
$R_{[\chi^{\ast}]_{av}}$. In all main panels (a) - (c) a double
logarithmic scale is considered. Additionally, all fitting
attempts are performed in the complete lattice range $L=20-200$
since no deviation in the estimated values of the critical
exponents was observed when shifting the value of the lower cutoff
$L_{min}$.}
\end{figure}
Thus, summarizing our estimates for the critical exponent $\nu$,
we feel that it is clear that it maintains the value $\nu=1$ of
the pure case, indicating again the validity of the strong
universality scenario.

We continue the presentation of our results by showing in
Fig.~\ref{fig:7} the FSS of the specific heat maxima averaged over
disorder: $[C]_{av}^{\ast}$ (up filled triangles) and
$[C^{\ast}]_{av}$ (down open triangles). Using these data for the
larger sizes $L\geq 60$, we tried to observe the quality of the
fittings, assuming a double-logarithmic divergence of the form
\begin{equation}
\label{eq:9} [C]_{av}^{\ast} ; [C^{\ast}]_{av}\sim
C_{1}+C_{2}\ln{(\ln L)},
\end{equation}
or a simple power law
\begin{equation}
\label{eq:10} [C]_{av}^{\ast} ; [C^{\ast}]_{av}\sim
C_{\infty}+C_{3}L^{\alpha/\nu}.
\end{equation}
Although it is rather difficult to numerically
distinguish between the above scenarios, our detailed fitting
attempts indicated that the double-logarithmic scenario
[Eq.~(\ref{eq:9})] applies better to the numerical data and this
is generally true for both $[C]_{av}^{\ast}$ and $[C^{\ast}]_{av}$ data.

In fact, the double-logarithmic fitting is shown in the main
panel, whereas in the corresponding inset of Fig.~\ref{fig:7} the
data of $[C^{\ast}]_{av}$ are plotted as a function of $\ln{(\ln
{L})}$. The solid line shown is an excellent linear fit for $L\geq
60$. Let us now give some details on the quality of the applied fittings. We used
the following sets of data
points $(L_{min}-L_{max})$, with $L_{max}=200$ and
$L_{min}=20,40,60,80$, and $100$. The quality of the fittings indicated a
very good trend for the values of $\chi^{2}$/DoF for the double logarithmic
fittings~(\ref{eq:9}) in the range: $0.2-0.7$ and for both sets of
data points. However, a strong reliability test in favor of the logarithmic
corrections scenario is provided by the stability of the
coefficient $C_{2}$, for both $[C]^{\ast}_{av}$
($C_{2}\approx 1.43(5)$) and $[C^{\ast}]_{av}$ ($C_{2}\approx
1.48(4)$) data. On the other hand, the estimated values of the
exponent $\alpha/\nu$ of the power law~(\ref{eq:10}), for both
$[C]_{av}^{\ast}$ and $[C^{\ast}]_{av}$, fluctuate in the range
$[-0.12(9),-0.05(6)]$ (as $L_{min}$ increases) with the
fitting procedure
becoming rather unstable as we move to larger values of
$L_{min}$. The conclusion is that our numerical data are
more properly described by the double logarithmic
form~(\ref{eq:9}), in agreement with the MC findings of Selke
\emph{et al.}~\cite{SSV-98} and Ballesteros \emph{et
al.}~\cite{BFMMPR-97} for the site-diluted SqIM and also with
those of Wang \emph{et al.}~\cite{wang-90} for the strong disorder
regime ($r=1/4$ and $r=1/10$) of the RBSqIM.

In Fig.~\ref{fig:8} we provide estimates for the magnetic
exponent ratios $\beta/\nu$ and $\gamma/\nu$ of the RBTrIM. In
panel (a) we plot the average magnetization at the estimated
critical temperature, as a function of the lattice size $L$ in a
log-log scale. The solid line is a linear fitting for $L\geq 20$
giving within error bars the value of the pure model, i.e.
$\beta/\nu=0.1253(5)\approx 0.125$. Additional estimate for the
ratio $\beta/\nu$ can be obtained from the FSS of the derivative
of the absolute order parameter with respect to inverse
temperature defined in Eq.~(\ref{eq:4}) which is expected to scale
as $L^{(1-\beta)/\nu}$ with the system size~\cite{ferrenberg-91}.
Thus, in panel (b) of Fig.~\ref{fig:8} we plot the data for
$\partial \langle |M|\rangle/\partial K$ averaged over disorder as
a function of $L$, also in a double-logarithmic scale. The solid
line is a linear fitting for the larger lattice sizes $L\geq 60$,
which combined with the value $\nu=1$, gives an estimate of
$0.1247(4)$ for the ratio $\beta/\nu$. Finally, in panel (c) we
present the FSS of the maxima of the average magnetic
susceptibility $[\chi]_{av}^{\ast}$ (filled squares) and also the
average of the individual maxima $[\chi^{\ast}]_{av}$ (open
triangles). The solid and dotted lines present linear fittings
using the total lattice range spectrum, giving the estimates
$1.751(5)$ and $1.756(9)$ for the ratio $\gamma/\nu$ in very good
agreement with the expected value $1.75$ of the pure system. For
the average $[\chi]_{av}^{\ast}$ the error bars indicate the
statistical errors due to the finite number of the realizations,
as discussed in Sec.~\ref{sec:1}. For the average
$[\chi^{\ast}]_{av}$ the errors bars shown reflect now the
relatively large sample-to-sample fluctuations.

Finally, using the latter
sample-to-sample fluctuations, we construct the ratio
$R_{[\chi^{\ast}]_{av}}$ and plot it as a function of the inverse
linear size, as shown in the inset of Fig.~\ref{fig:8}(c).
Clearly, for the present model the limiting value of
$R_{[\chi^{\ast}]_{av}}$ is non-zero, indicating, as expected also
for marginal disordered systems~\cite{wiseman-98}, a strong
violation of self-averaging.

\section{Conclusions}
\label{sec:5}

The effects induced by the presence of quenched bond randomness on
the critical behavior of the 2d Ising spin model embedded in the
triangular lattice have been investigated by an efficient
entropic scheme based on the Wang-Landau algorithm. In the first
part of our study we presented the finite-size scaling behavior of
the pure model, for which we calculated with high accuracy the
critical exponents and the coefficients of the
specific heat's logarithmic expansion at the critical point.
Our results are in full agreement with the exact
expansion presented by Wu \emph{et al.}~\cite{wu-03}.

In the second part of our study we investigated the critical
properties of the disordered system. The presented detailed
finite-size scaling analysis along the lines of the two existing
scenarios - strong versus weak universality - strongly supports
the scenario of strong universality. Thus, our results are in
agreement with the behavior predicted originally on theoretical
basis many years ago by Dotsenko and Dotsenko~\cite{DD-81},
Jug~\cite{jug-83}, Shalaev~\cite{shalaev-84},
Shankar~\cite{shankar-87}, and Ludwig~\cite{ludwig-87} and
verified by simulations in recent years for the square random
Ising model by several
authors~\cite{BFMMPR-97,AQS-97,SSLI-97,SSV-98,AQS-99,TO-01,kenna-08,fytas-08a}.
Particular interest was paid to the sample-to-sample fluctuations
of the random model and their scaling behavior that were used as a
successful alternative approach to estimate the critical
temperature and the correlation length's exponent. Closing, we
would like to note that another interesting candidate, that has
not been studied before in the triangular lattice, is the
$3$-state Potts model, which, in its pure version, has a positive
specific-heat exponent. Disorder would be relevant in this case
and could provide a further complementary study of the present
work, analogous to the early transfer-matrix calculations of
Derrida \emph{et al.}~\cite{derrida-87} of the random-bond model
on the square lattice.

\begin{acknowledgments}
The authors would like to thank Ralph Kenna and Walter Selke for
their useful comments and a critical reading of the manuscript.
\end{acknowledgments}

{}
\end{document}